\documentclass[twocolumn]{autart}    

\usepackage{graphicx}          

\usepackage{graphicx}      
\usepackage{natbib}        

\newcommand{\newsec}[1]{ {\bf \noindent #1. \quad \hspace{-0.5cm}}}

\newcommand{\mR}{\mathbb{R}}

\newcommand{\Exp}{\mathrm{Exp}}
\newcommand{\Ln}{\mathrm{Ln}}

\newcommand{\diag}{\mathrm{diag}}

\newcommand{\lbf}{\langle\!\langle}
\newcommand{\rbf}{\rangle\!\rangle}

\newcommand{\grad}{\mathrm{grad}\,}
\newcommand{\divv}{\mathrm{div}\,}

\newcommand{\trs}{\mathrm{\textsc{t}}}

\usepackage{amsmath,amssymb}

\newtheorem{rmk}{Remark}
\newtheorem{prp}[thm]{Proposition}
\usepackage[all]{xy}
\usepackage{psfrag}

\begin{document}
\begin{frontmatter}

\vspace{-0.2cm}
\title{\vspace{-0.2cm} Hamiltonian Perspective on Compartmental Reaction-Diffusion Networks}

\date{December 12, 2013}

\thanks[footnoteinfo]{Research is funded by the Netherlands Organisation for Scientific Research.}

\vspace{-0.9cm}
\author[A]{Marko Seslija}\ead{M.Seslija@rug.nl}, 
\author[B]{Arjan van der Schaft}\ead{A.J.van.der.Schaft@rug.nl}, 
\author[A]{Jacquelien M.A. Scherpen}\ead{J.M.A.Scherpen@rug.nl}
\address[A]{Department of Discrete Technology and Production Automation, Faculty of Mathematics and Natural Sciences, University of Groningen, Nijenborgh 4, 9747 AG Groningen, The Netherlands}
\address[B]{Johann Bernoulli Institute for Mathematics and Computer Science, University of Groningen, Nijenborgh 9, 9747 AG Groningen, The Netherlands}\vspace{-0.2cm}

\begin{abstract}
Inspired by the recent developments in modeling and analysis of reaction networks, we provide a geometric formulation of the reversible reaction networks under the influence of diffusion. Using the graph knowledge of the underlying reaction network, the obtained reaction-diffusion system is a distributed-parameter port-Hamiltonian system on a compact spatial domain. Motivated by the need for computer based design, we offer a spatially consistent discretization of the PDE system and, in a systematic manner, recover a compartmental ODE model on a simplicial triangulation of the spatial domain. Exploring the properties of a balanced weighted Laplacian matrix of the reaction network and the Laplacian of the simplicial complex, we characterize the space of equilibrium points and provide a simple stability analysis on the state space modulo the space of equilibrium points. The paper rules out the possibility of the persistence of spatial patterns for the compartmental balanced reaction-diffusion networks.
\vspace{-0.3cm}
\end{abstract}
\vspace{-0.3cm}
\vspace{-0.3cm}

\end{frontmatter}

\section{Introduction}
Reaction-diffusion systems model the evolution of the constituents distributed in space under the influence of chemical reactions and diffusion \cite{Smoller,Temam}. These spatially distributed models are essential for the understanding of many important phenomena concerning the development of organisms, coordinated cell behavior, and pattern formation \cite{MurrayBio}. Guided by the models of reaction-diffusion equations, designing multicellular systems for pattern formation is one of the present research topics in synthetic biology, with application foreseen in tissue engineering, biomaterial fabrication and biosensing \cite{Basu}. 

\vspace{-0.2cm}
The mathematical model of a chemical reaction that proceeds in an ideally diluted environment is a dynamical system $\dot x =f(x)$. A physicochemical interpretation would be that of a reaction system involving $m$ species $x_1,\ldots,x_m$ under the influence of kinetics modeled by the vector field $f$. If a well-mixed hypothesis is not reasonable, a more appropriate model is that of reaction-diffusion equations\vspace{-0.2cm}
\begin{equation}\label{eqRD}
\begin{split}
    \frac{\partial x}{\partial t}=\divv\!\!\left( D(x)\grad x \right) +f(x)\,,
\end{split}\vspace{-0.1cm}
\end{equation}
with
$x:=(x_1(\xi,t),\ldots,x_m(\xi,t))^\trs:(M,\mathbb{R}_{+})\rightarrow
\mathbb{R}^m$, and a positive semi-definite diagonal diffusion matrix. The operators
$\grad$ and $\divv$ act component-wise with respect to the local
coordinates $\xi=(\xi_1,\ldots, \xi_n)$ of the compact spatial domain $M\subset \mathbb{R}^n$. The constraints acting on the system
from the outside $M$ impose appropriate boundary conditions, which together with all the other technicalities will be discussed later.

\vspace{-0.2cm}
Motivated by the recent advances in the network control and graph theory, \cite{AJReaction} offers an elegant formulation for the dynamics of reversible chemical reactions \cite{HornJackson,Horn} and \cite{Feinberg,Feinberg1}. The graph description of the chemical reaction networks considered in \cite{AJReaction} has a direct thermodynamical interpretation, which can be regarded as a graph-theoretic version of the formulation derived in the work of \cite{op,opk}. Based on this formulation, \cite{AJReaction} characterizes the space of equilibrium points and provides a dynamical analysis on the state space modulo the space of equilibrium points.

\vspace{-0.2cm}
After a brief review of balanced reaction networks closely following the exposition of \cite{AJReaction}, we introduce a Dirac structure that captures the geometry of reaction-diffusion systems. We start from the fact that
the considered reaction systems are defined with respect to a finite Dirac structure on a manifold. This means that the reaction system from a network modeling perspective can be described by a set of energy-storing elements, a set of energy-dissipating (resistive) elements, and a set of ports (by which the interconnection is modeled), all interconnected by a power-conserving interconnections \cite{L2gain}. 

\vspace{-0.2cm}
From a control and interconnection viewpoint a prime desideratum is to formulate reaction diffusion systems with varying boundary conditions in order to allow energy flow through the boundary, since the interaction with the environment takes the place through the boundary. The Stokes-Dirac structure offers a geometric
framework for this and allows us to model balanced reaction-diffusion systems as distributed-parameter port-Hamiltonian systems.

\vspace{-0.2cm}
It is well-known that adding diffusion to the reaction system can generate behaviors absent in the ode case. This primarily pertains to the problem of diffusion-driven instability which constitutes the basis of Turing's mechanism for pattern formation \cite{Turing}, \cite{Prigogine}. Here, the port-Hamiltonian perspective permits us to
draw immediately some conclusions regarding passivity of reaction-diffusion systems, but also to claim the spatial uniformity of the asymptotic behavior of balanced reaction-diffusion system.

\vspace{-0.2cm}
In the second part of the paper, by adopting a discrete differential geometry-based approach and discretizing the reaction-diffusion system in port-Hamiltonian form, apart from preserving a geometric structure, a compartmental model analogous to the standard one is obtained \cite{Jacquez,Jovanovic}. Furthermore, we show the asymptotic stability of compartmental model and verify this result on an example of glycolisis pathway reactions.

\newsec{Notation}~The space of ${n}$ dimensional real vectors consisting of all strictly positive entries is denoted by $\mR_+^{n}$ and the space of ${n}$ dimensional real vectors consisting of all nonnegative entries is denoted by $\bar{\mR}_+^{n}$. ${\mathbf{1}}_m$ denotes a vector of dimension $m$ with all entries equal to 1. The time-derivative $\frac{dx}{dt}(t)$ of a vector $x$ depending on time $t$ will be usually denoted by $\dot{x}$. Define the mapping
$\mathrm{Ln} : \mathbb{R}_+^m \to \mathbb{R}^m, \quad x \mapsto \mathrm{Ln}(x),$
as the mapping whose $i$-th component is given as
$\left(\mathrm{Ln}(x)\right)_i := \mathrm{ln}(x_i).$
Similarly, define the mapping
$\mathrm{Exp} : \mathbb{R}^m \to \mathbb{R}_+^m, \quad x \mapsto \mathrm{Exp}(x),$
as the mapping whose $i$-th component is given as
$\left(\mathrm{Exp}(x)\right)_i := \mathrm{exp}(x_i).$
Also, define for any vectors $x,z \in \mathbb{R}^m$ the vector $x \cdot z \in \mathbb{R}^m$ as the element-wise product $\left(x \cdot z\right)_i :=x_iz_i, \, i=1,2,\ldots,m,$ and the vector $\frac{x}{z} \in \mathbb{R}^m$ as the element-wise quotient $\left(\frac{x}{z}\right)_i := \frac{x_i}{z_i}, \, i=1,\cdots,m$. Note that with these notations $\Exp (x + z) = \Exp (x) \cdot \Exp (z)$ and $\Ln (x \cdot z) = \Ln (x) + \Ln (z),~ \Ln \left(\frac{x}{z}\right) = \Ln (x) - \Ln (z)$.
On an $n$-dimensional smooth manifold $M$, the space of smooth scalar valued functions will be denoted by $C^\infty(M)$, while the space of vector valued functions will be $C^\infty(M;\mathbb{R}^n)$. We employ the notion $C_m^\infty (M):=C^\infty (M)\times\cdots\times C^\infty (M)$, with the product being taken $m$ times.

\section{Balanced Chemical Reaction Networks}\label{Sec2}
\vspace{-0.3cm}
\newsec{Stoichiometry}~Consider a chemical reaction network involving $m$ chemical species (metabolites), among which $r$ chemical reactions take place. The basic structure underlying the dynamics of the vector $x \in \bar{\mathbb{R}}_+^m$ of concentrations $x_i, i=1,\dots, m,$ of the chemical species is given by the {\it balance laws}: $\dot{x} = Sv$, where $S$ is an $m \times r$ matrix, called the {\it \textbf{stoichiometric matrix}}. The elements of the vector $v \in \mathbb{R}^r$ are commonly called the (reaction) {\it fluxes}. The stoichiometric matrix $S$, which consists of (positive and negative) integer elements, captures the basic conservation laws of the reactions.

\vspace{-0.3cm}
\newsec{The Complex Graph}~The network structure of a chemical reaction network cannot be directly captured by a graph involving the chemical species, because, in general, there are more than two species involved in a reaction. Following the approach originating in the work of \cite{HornJackson,Horn} and \cite{Feinberg,Feinberg1}, we will introduce the space of {\it\textbf{complexes}}. The set of complexes of a chemical reaction network is simply defined as the union of all the different left- and right-hand sides (substrates and products) of the reactions in the network. The expression of the complexes in terms of the chemical species is formalized by an $m \times c$ matrix $Z$, whose $\rho$-th column captures the expression of the $\rho$-th complex in the $m$ chemical species.

\vspace{-0.3cm}
Since the complexes are left- and right-hand sides of the reactions they can be naturally associated with the vertices of a {\it directed graph}, with edges corresponding to the reactions. The resulting graph is called the {\it complex graph}. The complex graph is defined by its $c \times r$ {\it \textbf{incidence matrix}} $B$, $c$ being the
number of vertices and $r$ being the number of edges. There is a close relation between the matrix $Z$ and the stoichiometric matrix $S$, which is expressed as $
S = ZB$. For this reason we will call $Z$ the {\it \textbf{complex stoichiometric matrix}}. Hence, the relation $\dot{x} = Sv$ can be also written as $\dot{x} = ZBv$, with the vector $Bv$ belonging to the space of complexes $\mathbb{R}^c$.

\vspace{-0.3cm}
\newsec{Balanced Mass Action Kinetics}
A vector of concentrations $x^* \in \mathbb{R}^m_+$ is called an {\it equilibrium} for the dynamics $\dot{x} =Sv(x)$ if 
$
Sv(x^*) = 0$. Furthermore, $x^* \in \mathbb{R}^m_+$ is called a {\emph{\textbf{thermodynamic equilibrium}}} if $v(x^*)=0
$.
Clearly, any thermodynamic equilibrium is an equilibrium, but not necessarily the other way around (since in general $S=ZB$ is not injective). 

\vspace{-0.2cm}
Consider the $j$-th reaction from substrate $\mathcal{S}_{j}$ to product $\mathcal{P}_{j}$, described by the mass action rate equation\vspace{-0.2cm}
\[
v_{j}(x) = k^{\text{forw}}_{j} \exp\big(Z_{\mathcal{S}_{j}}^\textsc{t} \mathrm{Ln}(x)\big)-k^{\text{rev}}_{j} \exp\big(Z_{\mathcal{P}_{j}}^\textsc{t} \mathrm{Ln}(x)\big),
\]

\vspace{-0.65cm}
where $Z_{\mathcal{S}_j}$ and $Z_{\mathcal{P}_j}$ denote the columns of the complex stoichiometry matrix $Z$ corresponding to the substrate and the product complexes of the $j$-th reaction. Then $x^* \in \mathbb{R}^m_+$ is a thermodynamic equilibrium, i.e., $v(x^*)=0$, if and only if $\kappa_{j}(x^*) :=k_{j}^{\text{forw}} \exp \left(Z_{\mathcal{S}_{j}}^\textsc{t} \Ln (x^*)\right)=k_{j}^{\text{rev}} \exp \left(Z_{\mathcal{P}_{j}}^\textsc{t} \Ln (x^*)\right)$, for $ j = 1, \cdots, r.$ The mass action reaction rate of the $j$-th reaction now can be written as
\[
v(x) = - \mathcal{K}(x^*) B^\textsc{t} \mathrm{Exp} \left(Z^\textsc{t} \mathrm{Ln}\left(\frac{x}{x^*}\right)\right),
\]
where $\mathcal{K}(x^*) $ is the $r \times r$ is a positive diagonal matrix of balanced reaction constants given as $\mathcal{K}(x^*) := \mathrm{diag} \big( \kappa_1(x^*), \cdots, \kappa_r(x^*) \big)$.

The dynamics of a \textbf{\emph{balanced reaction network}} takes the form
\begin{equation}\label{masterequation}
\dot{x} = - Z B \mathcal{K}(x^*) B^\textsc{t} \mathrm{Exp} \left(Z^\textsc{t} \mathrm{Ln}\left(\frac{x}{x^*}\right)\right). 
\end{equation}
This form will be the starting point for the analysis of balanced chemical reaction networks in the rest of this paper. Furthermore, we shall assume the validity of the \emph{global persistency conjecture,} which states that for a positive initial condition $x_0\in \mathbb{R}_{+}^m$, the solution $x$ of (\ref{masterequation}) satisfies: $\mathrm{lim}\,\mathrm{inf}_{t\rightarrow \infty}x(t)>0$. The global persistency conjecture recently was proven for the single linkage class case in \cite{Anderson}, but for the system (\ref{masterequation}) remains an open problem.

\vspace{-0.3cm}
\newsec{Stability of Balanced Reaction Networks}~It follows that once a thermodynamic equilibrium $x^*$ is given, the set of {\it all} thermodynamic equilibria is described by the following proposition.

\vspace{-0.2cm}
\begin{prp}[\cite{AJReaction}]\label{thermoequilibrium1}
Let $x^* \in \mathbb{R}^m_+$ be a thermodynamic equilibrium, then the set of {\it all} thermodynamic equilibria is given by 
\begin{equation}\label{equilibria}
\mathcal{E} := \{ x^{**} \in \mathbb{R}^m_+ \mid S^\textsc{t} \Ln \left(x^{**}\right) = S^\textsc{t} \Ln \left(x^{*}\right) \}.
\end{equation}
\end{prp}

\vspace{-0.3cm}
Making use of the formulation of the dynamics of balanced reaction networks in (\ref{masterequation}), in \cite{AJReaction} it was shown that all equilibria of a balanced reaction network are actually thermodynamic equilibria, and thus given by (\ref{equilibria}). A similar result was obtained in the classical papers \cite{Horn,HornJackson,Feinberg1}, but also in \cite{Sontag}, for a different class of chemical reaction networks.%

\vspace{-0.2cm}
\begin{thm}[\cite{AJReaction}]\label{tm:Lyap}
Consider a balanced mass action reaction network given by (\ref{masterequation}), and in addition assume the global persistency property, then for every initial condition $x(0)\in \mR_+^{m}$, the species concentration $x$ converges for $t \to \infty$ to $\mathcal{E}$.
\end{thm}

\vspace{-0.2cm}
In the remaining of the paper, we will extend some of the results of balanced chemical reaction networks to spatially distributed systems.

\section{Geometric Formulation}
In classical field theories the geometric content of the physical variables is usually expressed by identifying them with differential forms of appropriate order \cite{vdSM02}. Previously in \cite{SeslijaRD} we have offered a Stokes-Dirac structure for reaction-diffusion systems. In this paper we avoid the exterior formulation and introduce a Dirac structure for reaction-diffusion systems defined on the space of smooth functions on a Riemannian manifold with boundary.

\begin{figure}
\centering
   \includegraphics[width=5.8cm,angle=270]{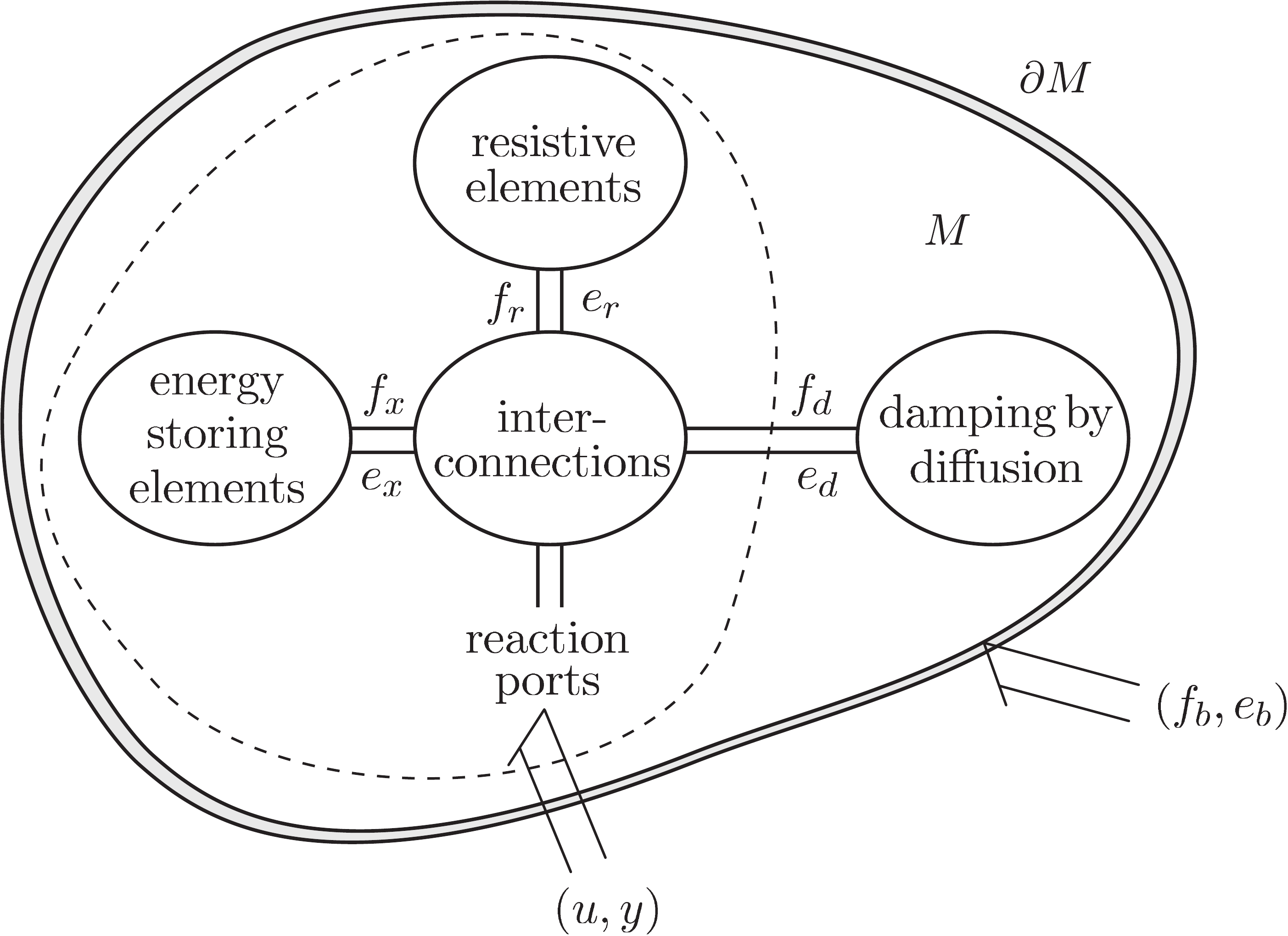}\\
  \caption{Reaction-diffusion system as a dissipative distributed port-Hamiltonian system. The conjugate variables $u$ and $y$ represent the inflows and the outflows of the reaction dynamics. In this paper the reaction system is considered to be closed; that is, either $u=0$ or $y=0$.}\label{fig:ChRDOnManifold}
\end{figure}

\vspace{-0.4cm}
The formulation of a reaction-diffusion system as a port-Hamiltonian
system on a \emph{compact} $n$-dimensional smooth \emph{Riemannian} manifold $M$ with boundary $\partial M$ is given as follows. We identify the mass density variables
with an $m$ component vector of scalar valued functions, that is $x\in
C_m^\infty(M)$. The influence of
the external world (reaction-diffusion system) to the system
(outside world) is modeled through the boundary efforts $e_b\in
C_m^\infty(\partial M)$ (and boundary flows $f_b\in
C_m^{\infty}(\partial M)$). The reaction part is in its
nature finite-dimensional and as such is modeled as the
interconnection of the atomic elements, each of them characterized
by a particular energetic behavior (energy storing, energy
conversion or dissipation). Each of these elements can interact with
the environment by means of a port---a couple of inputs and outputs
whose combination gives the power flow. The transport of
the constituents in space is governed by the laws of diffusion,
which is modeled as a thermal damping by termination of the
appropriate ports (see Figure~\ref{fig:ChRDOnManifold}).

\vspace{-0.4cm}
Let the space of flows be $\mathcal{F}$ and its dual the space of efforts be $\mathcal{E}$. We set $\mathcal{F}:=\mathcal{F}_x\oplus \mathcal{F}_d\oplus \mathcal{F}_r\oplus \mathcal{F}_b$ and $\mathcal{E}:=\mathcal{E}_x\oplus \mathcal{E}_d\oplus \mathcal{E}_r\oplus \mathcal{E}_b$, where $\mathcal{F}_x=\mathcal{E}_x=C_m^\infty (M)$ is the carrier space of concentrations, $\mathcal{F}_d=\mathcal{E}_d=C_m^\infty (M;\mathbb{R}^n)$ the space of gradients, $\mathcal{F}_r=\mathcal{E}_r=C_m^\infty (M)$ the reaction flows, $\mathcal{F}_b=C_m^\infty (\partial M)$ being the space of boundary fluxes, and $\mathcal{E}_b=C_m^\infty (\partial M)$ be the space of chemical potentials restricted to the boundary $\partial M$.

The inner product in $C_m^\infty(M)$ is given by
\[
\left\langle \alpha, \beta\right\rangle_{L_m^2(M) }=\int_M \alpha^\textsc{t}(\xi)  \beta(\xi) \mathrm{d}\xi , \quad\alpha,\,\beta\in C_m^\infty(M),
\]
where $\mathrm{d}\xi:=\mathrm{d}\xi_1 \cdots\mathrm{d}\xi_n$ is the volume element on $M$. Similarly, the inner product on the boundary $\partial M$ is defined by\vspace{-0.3cm}
\[
\left\langle \alpha, \beta\right\rangle_{L_m^2 (\partial M)}=\int_{\partial M} \alpha^\textsc{t}(\xi)  \beta(\xi) \mathrm{d}\xi_A, \quad\alpha,\,\beta\in C_m^\infty(\partial M),
\]
where $\mathrm{d}\xi_A:=\mathrm{d}\xi_1\cdots\mathrm{d}\xi_{n-1}$ is the volume form on the boundary $\partial M$. Analogously, we defined the inner product in $C_m^\infty(M; \mathbb{R}^n)$ and $C_m^\infty(\partial M; \mathbb{R}^{n-1})$.

\vspace{-0.3cm}
A non-degenerate pairing between $\mathcal{F}$ and $\mathcal{E}$ is
defined by the following bilinear form on $\mathcal{F}\times
\mathcal{E}$ with values in $\mathbb{R}$
\begin{equation*}\label{eqFERD}
\begin{split}
    \lbf
    &(f_x^1,f_d^1,f_\gamma^1,f_b^1, e_x^1, e_d^1,e_r^1,e_b^1),(f_x^2,f_d^2,f_r^2,f_b^2,e_x^2,e_d^2,e_r^2,e_b^2)\rbf\\
    &:= \left\langle f_x^1, e_x^2\right\rangle_{L_m^2(M)}+ \left\langle f_d^1, e_d^2\right\rangle_{L_m^2(M)} + \left\langle f_r^1, e_r^2\right\rangle_{L_m^2(M)}\\
    &\qquad+ \left\langle f_x^2, e_x^1\right\rangle_{L_m^2(M)}+ \left\langle f_d^2, e_d^1 \right\rangle_{L_m^2(M)}+ \left\langle f_r^2, e_r^1\right\rangle_{L_m^2(M)}\\
    &\qquad+ \left\langle f_b^1, e_b^2\right\rangle_{L_m^2(\partial M)}+  \left\langle f_b^2, e_b^1\right\rangle_{L_m^2(\partial M)},
\end{split}
\end{equation*}
where $(f_x^i,f_d^i, f_r^i, f_b^i)\in \mathcal{F}$ and $(e_x^i,e_d^i, e_r^i, e_b^i)\in \mathcal{E}$, $i=1,2$. 

\vspace{-0.3cm}
The Stokes-Dirac structure that underpins the geometry of reaction-diffusion systems is a maximally isotropic subspace of $\mathcal{F}\times \mathcal{E}$. The following theorem gives the construction of a such Dirac structure.

\begin{thm}\label{Th:DiracStructure}
Define $\mathcal{D} \subset \mathcal{F}\times\mathcal{E}$ by
\begin{equation}
\begin{split}
\mathcal{D}:=\big\{  (f_x,f_d,&f_r,f_b,e_x,e_d,e_r,e_b) \in  \mathcal{F}\times\mathcal{E}
    \,\big|\\
\left(\begin{array}{c}f_x \\ f_d \\ e_r \end{array}\right)
&=\left(\begin{array}{ccc}0 &  \mathrm{div} & -Z \\
 \mathrm{grad}  & 0 & 0 \\
 Z^\textsc{t} & 0 & 0\end{array}\right)\left(\begin{array}{c}e_x \\e_d \\f_r\end{array}\right),\\
\left(\begin{array}{c}e_b \\f_b\end{array}\right)
&=\left(\begin{array}{ccc}\mathrm{tr} & 0 & 0 \\0 &  \nu\!\cdot\!\mathrm{tr} & 0\end{array}\right)\left(\begin{array}{c}e_x \\e_d \\f_r\end{array}\right)\big\},
\end{split}
\end{equation}
where $Z$ is an $m\times c$ matrix, $\mathrm{tr}$ is the trace on the boundary $\partial M$ and $\nu$ is the unit normal on $\partial M$. The subbundle $\mathcal{D}$ is a Dirac structure with respect to the bilinear form $\lbf,\rbf$.
\end{thm}
\vspace{-0.5cm}
\begin{pf}
Using the fact that $\langle Z f_r,e_x\rangle_{L_m^2(M)}=\langle f_r, Z^\textsc{t} e_x \rangle_{L_m^2(M)}$ for any $f_r\in C_m^\infty(M)$ and $e_x\in C_m^\infty(M)$, and applying the integration by parts formula
\begin{equation*}
\begin{split}
\langle {\mathrm{grad}}\,e_x,e_d\rangle_{L_m^2(M)} =& \,\langle e_x,\mathrm{div}\, e_d\rangle_{L_m^2(M)}
\\
&+\langle
\mathrm{tr}\,e_x,\mathrm{tr}\,e_d \cdot\nu\rangle_{L_m^2(\partial M)},
\end{split}
\end{equation*}
for $e_x\in C_m^{\infty}(M),\ e_d\in C_m^\infty(M)$, similar to Theorem~1 in  \cite{vdSM02}, it is routine to show that $\mathcal{D=D}^\perp$. \hfill $\qed$
\end{pf}

\section{Reaction-Diffusion Dynamics}

In order to obtain a port-Hamiltonian formulation of
reaction-diffusion systems, we start from the Gibb's free energy associated to the reaction system given by
\begin{equation}\label{eq:FreeEnergyGibbs}
G(x) =x^\textsc{t} \mathrm{Ln}\left(\frac{x}{x^*}\right) +  \left(x^* - x \right)^\textsc{t} {\mathbf{1}}_m,
\end{equation}
where $x^*$ is an equilibrium of the reaction network and $\mathbf{{1}}_m$ denotes a vector of dimension $m$ with all ones. It can be immediately checked that the gradient of $G$ is the chemical potential $\mu$, i.e., $\frac{\partial G}{\partial x}(x) = \mathrm{Ln}\left( \frac{x}{x^*}\right) =:\mu(x)$. The energy, the \emph{Hamiltonian}, of the reaction-diffusion system is
$\mathcal{G}=\int_M G \mathrm{d }\xi$.

Starting from the Dirac structure $\mathcal{D}$ given in Theorem~\ref{Th:DiracStructure}, define the energy storage relations
\begin{equation}\label{eq:pHenergy}
\begin{split}
  f_x=-\frac{\partial x}{\partial t}\,,~~~~e_x=\frac{\partial G}{\partial x}(x)\,.
\end{split}
\end{equation}

The power-dissipation of the reaction part is
\begin{equation}\label{eq:pHdissipation}
\begin{split}
 f_r=-B \mathcal{K}(x^*) B^\textsc{t}\mathrm{Exp}(e_r)\,,
\end{split}
\end{equation}
where the mapping $B \mathcal{K}(x^*) B^\textsc{t}\mathrm{Exp}(\cdot)$ satisfies
\begin{equation}\label{eq:pHdissipationP}
\begin{split}
 e_r^\textsc{t} f_r \leq 0~~\textrm{for~all}~~ e_r\in \mathcal{E}_r\,.
\end{split}
\end{equation}
This is due to the fact that the exponential function is strictly increasing, so
\begin{equation*}
\begin{split}
\gamma^\textsc{t} B\mathcal{K} B^\textsc{t} \mathrm{Exp}(\gamma)&= \sum_{j=1}^r
\left(\gamma_{\mathcal{P}_j}(x) - \gamma_{\mathcal{S}_j}(x)\right)\kappa_j(x^*)\\&~~~~~~~~~~\cdot \left(\exp\big(\gamma_{\mathcal{P}_j}(x) \big)-\exp\big(\gamma_{\mathcal{S}_j}(x) \big)\right)\\
& \geq 0
\end{split}
\end{equation*}
for $\kappa_j(x^*) >0$, $j=1,\ldots, r$.

Ordinarily, diffusion is treated as power-dissipation by thermal
motion of particles and as such quantitatively is characterized by
the diffusion matrix $D$ introduced in (\ref{eqRD}), which
explicitly refers to the state variable $x$. In the work at hand
diffusion is modeled by termination of the diffusion port as
\begin{equation}\label{eq:pHdiffusion}
\begin{split}
  e_d=-\mathcal{R}_d(f_d)\,,
\end{split}
\end{equation}
where $\mathcal{R}_d:\mathcal{F}_d\rightarrow
\mathcal{E}_d$ is in general a nonlinear mapping satisfying dissipation inequality
\[
 e_d^\textsc{t} f_d \leq 0~~\textrm{for~all}~~ f_d\in \mathcal{F}_d\,.
\]
The operator $\mathcal{R}_d$, instead of acting upon the gradient of the state $x$, for the argument takes the gradient of the co-energy variables (i.e., the chemical potential $\mu$). We call this operator the energy
diffusion operator.

When the diffusion operator $\mathcal{R}_d$ is a matrix function of the state $x$, these constitutive relations define the reaction-diffusion system in the port-Hamiltonian framework as
\begin{equation}\label{eq:pHRDvec-calRel3Rd}\vspace{-0.3cm}
\begin{split}
  \frac{\partial x}{\partial t}&=\divv \!\!\left(\mathcal{R}_d(x)\grad  \frac{\partial G}{\partial x}(x) \right)
  \\
  &~~~- Z B \mathcal{K}(x^*) B^\textsc{t} \mathrm{Exp} \left(Z^\textsc{t}\frac{\partial G}{ \partial x}(x)\right)\\
  e_b&=\frac{\partial G}{\partial x}(x)  |_{\partial M}\\
  f_b&=\mathcal{R}_d(x) \grad\frac{\partial G}{\partial x}(x) \cdot \nu|_{\partial M}\,,
\end{split}\vspace{-0.3cm}
\end{equation}
with the Hamiltonian is $\mathcal{G}=\int_M G \mathrm{d }\xi$ and $G$ given by (\ref{eq:FreeEnergyGibbs}).

Because $\grad \mathrm{Ln}\left( \frac{x}{x^*} \right)=\mathrm{diag}\left(\frac{1}{x_1}, \dots, \frac{1}{x_m}\right) \mathrm{grad}\,x$, the system (\ref{eq:pHRDvec-calRel3Rd}) is in the form (\ref{eqRD}) with $\mathcal{R}_d(x)=\mathrm{diag}\left({x_1}, \dots, {x_m}\right) D(x)$ and the reaction dynamics $f(x)=-ZB \mathcal{K}(x^*) B^\textsc{t}\mathrm{Exp}\left(Z^\textsc{t}\mathrm{Ln}\left( \frac{x}{x^*} \right)\right)$.

\newsec{Standard Model}~The dynamical analysis of the balanced reaction networks presented in \cite{AJReaction} is given on the state space modulo the space of equilibrium points. For the sake of thermodynamical consistency, we rewrite the system (\ref{eq:pHRDvec-calRel3Rd}) into the form given in terms of the disagreement vector $\frac{x}{x^*}$ as
\begin{equation}\label{eq:pHRDvec-calRel3RdNew}\vspace{-0.3cm}
\begin{split}
  \frac{\partial x}{\partial t}&=\divv \!\!\left(R_d(x)\grad\!\!\left( \frac{x}{x^*}\right) \right)+f(x)\\
  e_b&=\mathrm{Ln}\left( \frac{x}{x^*} \right) |_{\partial M}\\
  f_b&=R_d(x)\grad\! \!\left(\frac{x}{x^*}\right)\cdot \nu|_{\partial M}\,,
\end{split}\vspace{-0.2cm}
\end{equation}
where $R_d(x):=\mathcal{R}_d(x) \mathrm{diag}\left(\frac{x_1^*}{x_1}, \dots, \frac{x_m^*}{x_m}\right)$ and $f$ is given by the right-hand side of (\ref{masterequation}) .

The existence of solutions for the systems (\ref{eq:pHRDvec-calRel3Rd}) and (\ref{eq:pHRDvec-calRel3RdNew}) is a complex issue. The papers \cite{Morgan,Fitzgibbon} do provide a working framework for the systems with separable Lyapunov functions. Furthermore, according to \cite{Fitzgibbon}, the system \emph{does not }generate spatial patterns. The problem of the existence of classical solution and spatial uniformity of the steady state in the presence of Neumann's boundary conditions for the semilinear reaction-diffusion systems we plan to address in a separate contribution.


\newsec{Passivity}
Define the complex affinity as $\gamma(x)=Z^{\textsc{t}}\Ln\left(\frac{x}{x^*}\right)$. Assuming the existence of a classical solution to (\ref{eq:pHRDvec-calRel3Rd}), as an immediate consequence we obtain the following energy balance
\begin{equation*}\label{eq:CHRD:timederG}
\begin{split}
\frac{d}{dt} \mathcal{G}(x) &= \left\langle \frac{\partial G}{\partial x}(x), \frac{\partial x}{\partial t}\right\rangle_{L_m^2(M)}=\left\langle \mu(x), \frac{\partial x}{\partial t}\right\rangle_{L_m^2(M)}\\
&=-\left\langle Z^\textsc{t}\mu(x), B \mathcal{K}(x^*) B^\textsc{t} \mathrm{Exp}(Z^\textsc{t}\mu(x))\right\rangle_{L_m^2(M)}\\
&~~~+\left\langle\mu(x), \divv \!\!\left(\mathcal{R}_d(x)\grad \mu(x)\right)\right\rangle_{L_m^2(M)}\\
&=-\left\langle\gamma(x), B \mathcal{K}(x^*) B^\textsc{t} \mathrm{Exp}(\gamma(x))\right\rangle_{L_m^2(M)}\\
&~~~-\left\langle\grad\mu(x), \mathcal{R}_d(x)\grad \mu(x) \right\rangle_{L_m^2(M)}\\
&~~~+\left\langle\mu(x), \mathcal{R}_d(x)\grad \mu(x)\cdot \nu \right\rangle_{L_m^2(\partial M)}.
\end{split}
\end{equation*}
Because the exponential function is strictly increasing the following inequality holds
\begin{equation*}
\begin{split}
\gamma^\textsc{t} B\mathcal{K} B^\textsc{t} \mathrm{Exp}(\gamma)
 \geq 0
\end{split}
\end{equation*}
for $\kappa_j(x^*) >0$, $j=1,\ldots, r$, which immediately implies
$$\left\langle Z^\textsc{t}\mu(x), B \mathcal{K}(x^*) B^\textsc{t} \mathrm{Exp}(Z^\textsc{t}\mu(x))\right\rangle_{L_m^2(M)} \geq 0\,. $$
Furthermore, since
$$\left\langle\grad\mu(x), \mathcal{R}_d(x)\grad \mu(x) \right\rangle_{L_m^2(M)}\geq 0,$$ the {\it passivity} property holds
\begin{equation}
\frac{d}{dt} \mathcal{G} \leq \left\langle e_b, f_b \right\rangle_{L_m^2(\partial M)},
\end{equation}
which means that the Hamiltonian functional $\mathcal{G}$ is non-increasing along solution trajectories of (\ref{eq:pHRDvec-calRel3Rd}). The equal conclusion, of course, also holds for the system (\ref{eq:pHRDvec-calRel3RdNew}).


\section{Structure-Preserving Discretization}\label{Ch:RD:SecStru-Pres}

\newsec{A Single Species System}
Firstly, let us consider the single component reaction-diffusion system
\begin{equation}\label{eq:RDsimple1com}
\begin{split}
\frac{\partial x}{\partial t}&= \mathrm{div} \left( D(x) \,\mathrm{grad}\, x\right)+g(x)\\
e_b&=x|_{\partial M}\\
f_b&= \mathrm{grad}\,x\cdot \nu|_{\partial M},
\end{split}
\end{equation}
where $x, g, D\in C^\infty(M)$, $D(x)>0$ for all $x$, and $e_b\in C^\infty(\partial M)$ and $f_b\in C^\infty(\partial M)$.

In the framework of exterior geometry, we can identify $0$- and $n$-forms with scalar valued functions, while $1$-  and $(n-1)$-forms are identified with proxy fields. This allows for the operators $\mathrm{grad}$ and $\mathrm{div}$ to be rewritten in terms of the exterior derivative $\mathrm{d}$ and the Hodge star $*$ \emph{formally} as: $\mathrm{grad}=\mathrm{d}$ and $\mathrm{div}=-*\mathrm{d}*$; for more details see, e.g., \cite{Arnold}. 

Following the exposition of \cite{SeslijaAutomatica}, let $K$ be a homological simplicial complex obtained by triangulation of the manifold $M$. Assuming that $K$ is well-centered, its circumcentric dual is $\star K=\star_\mathrm{i} K\times \star_\mathrm{b} K$, where $\star_\mathrm{i} K$ is the interior dual and $\star_\mathrm{b} K$ is the boundary dual, as explained in \cite{SeslijaAutomatica,SeslijaJGP}.

The discrete analogue of an oriented manifold is an oriented simplicial complex, while differential forms are discretized as cochains. A $k$-cochain is a real-valued function on the $k$-simplices of $K$, which we will also call a \emph{discrete} $k$-form. Analogously, we define the space of discrete forms on $\star_\mathrm{i} K$ and $\star_\mathrm{b} K$. By $\Omega_d^k(K)$, $\Omega_d^k(\star_\mathrm{i} K)$, and $\Omega_d^k(\star_\mathrm{b} K)$ we denote the space of the primal $k$-cochains, the dual $k$-cochains, and the boundary dual $k$-cochains, respectively.

The discrete analogue of (\ref{eq:RDsimple1com}) is
\begin{equation}\label{eq:RDsimple1comDiscrete}
\begin{split}
\!\!\frac{\partial x}{\partial t}&= \left(*_0\right)^{-1}\!\!\left(  {\mathbf{d}}_\mathrm{i}^{n-1} \!*_1\! D_d(x)\mathbf{d}^0 x + \mathbf{d}_\mathrm{b}^{n-1} \hat f_b\right)+g(x)\!\!\\
e_b&={\mathbf{tr}^0}\,x,
\end{split}
\end{equation}
where the state $x$ now lives on the set of verices of $K$, that is, $x\in \Omega_d^0(K)$, the input $\hat f_b\in \Omega_d^{n-1}(\star_\mathrm{b} K)$, and the output $e_b\in \Omega_d^{0}(K)$. The positive-definite (discrete) diffusion matrix is $x\mapsto D_d(x)\in \mathbb{R}^{N_e\times N_e}$, with $N_e=\mathrm{dim}\Omega_d^{1}(K)$, while the operators $*_0$, $*_1$, $\mathbf{d}^0$, $ {\mathbf{d}}_\mathrm{i}^{n-1} $, $\mathbf{d}_\mathrm{b}^{n-1}$, and ${\mathbf{tr}^0}$ have been defined in \cite{SeslijaAutomatica}. The operator $\mathbf{d}^0:\Omega(K)\rightarrow \Omega^1(K)$ is nothing but the transpose of the incidence matrix of the primal skeleton (from the primal edges to the primal vertices). Furthermore, $\mathbf{d}^0=-\left({\mathbf{d}}_\mathrm{i}^{n-1} \right)^\textsc{t}$ and $\mathbf{d}_\mathrm{b}^{n-1}=\left({\mathbf{tr}^0}\right)^\textsc{t}$. The discrete Hodge operator $*_1:\Omega^1(K)\rightarrow \Omega^{n-1}( \star_\mathrm{i} K)$ is a diagonal matrix with the $k$-th entry being equal $|\star_\mathrm{i} \sigma_k^1|/|\sigma_k^1|$, where $\sigma_k^1$ is the primal edge with the dual $\star_\mathrm{i} \sigma_k^1$. The matrix $*_0$ is a diagonal matrix whose $k$-th element is $|\star_\mathrm{i} \sigma_k^0|/|\sigma_k^0|$.

\begin{rmk}
In fact, the model (\ref{eq:RDsimple1comDiscrete}) slightly, but crucially, differs from the standard compartmental model on graphs, where the matrix $(*_0)^{-1}$ does not appear \cite{Arcak}. This implies that in the standard graph model $*_0=I_N$, $N={\mathrm{dim}}\,\Omega_d^0(K)$, that is, $|\star_\mathrm{i} \sigma_k^0|=1$ for all $k=1,\ldots, N$, meaning that all the compartments are of \emph{equal} volume. This fact does not surprise, since the graph formulation \emph{does not} capture the \emph{geometric} content of the underlying model.
\end{rmk}

\begin{figure}
\centering
\vspace{-0.0cm}
      \hspace{0cm}\includegraphics[width=7.0cm]{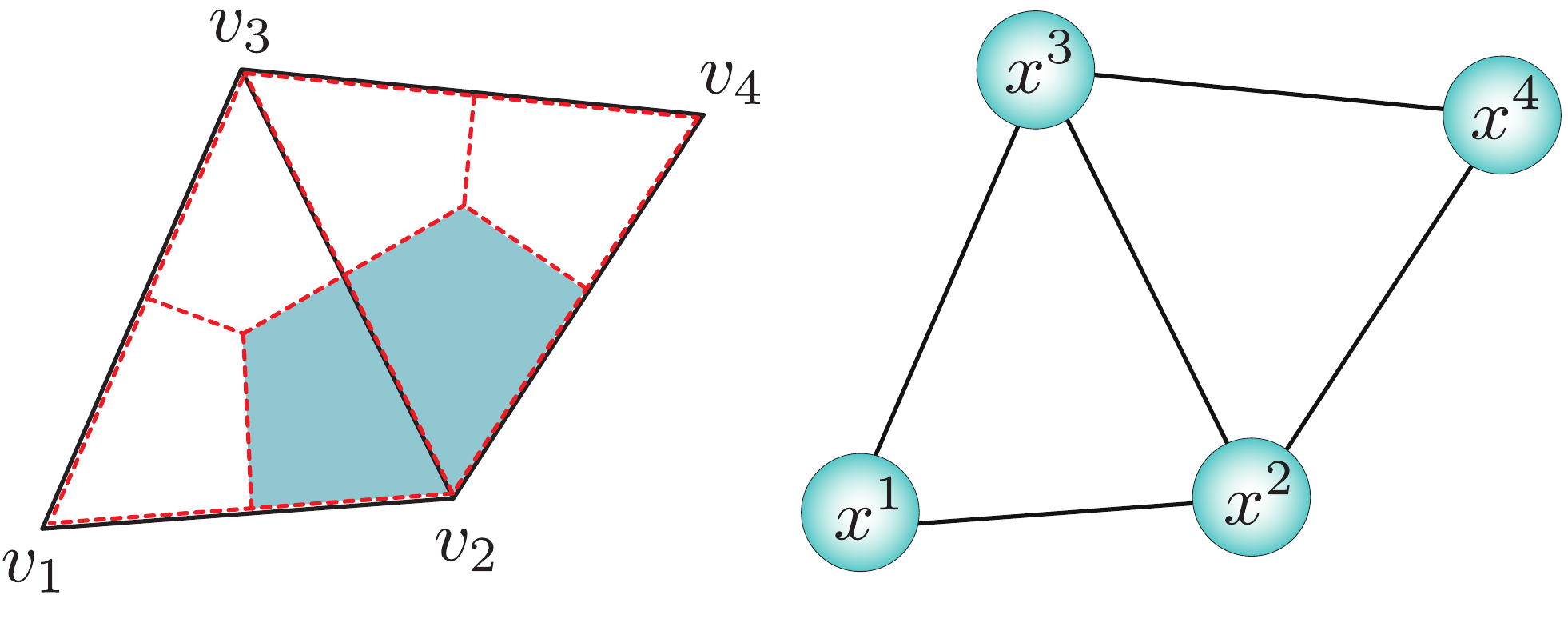}
      \vspace{-0.1cm}
  \caption{A simplicial complex $K$ consists of two triangles. The dual edges introduced by the circumcentric subdivision are shown dotted. The state vector $x^j=\left(x_1,\ldots, x_m \right)^{\textsc{t}}$ is associated to the vertex $v_j$ for each $j\in\{1,\ldots, N\}$. The number of compartments for this example is $N=4$. The shaded region, the dual cell $*_\mathrm{i} v_2$ of the vertex $v_2$, represents the compartment with the state $x^2$.}\label{fig:CH:RDcompart}
\end{figure}

\newsec{Multicomponent System}
Let us now consider the reaction-diffusion system with $m$ components (cf. \ref{eq:pHRDvec-calRel3Rd}). To each node of the primal mesh we associate reaction dynamics. That is, to a node $\sigma_j^0$ we associate the state $x^j\in {\mathbb{R}}_+^m$. The geometric dual of $\sigma_j^0$, $\star_\mathrm{i }\sigma_j^0$, is the dual volume cell which represents the $j$-th compartment (see Figure~\ref{fig:CH:RDcompart}). The number of the compartments is $N=\mathrm{dim}\,\Omega_d^0(K)=\mathrm{dim}\, \Omega_d^n(\star_\mathrm{i} K)$. The compartments interact with each other through the diffusion modeled as follows.

By $X$ denote the concatenated vector
\vspace{-0.2cm}
\begin{equation}
\label{eq:RDconcatenatedX}
X=\left( \left(x^1\right)^\textsc{t}, \dots, \left(x^N\right)^\textsc{t} \right)^\textsc{t},
\end{equation}

\vspace{-0.6cm}
where $x^j\in {\mathbb{R}}_+^m$, and let

\vspace{-0.6cm}
\begin{equation}
\label{eq:RDconcatenatedF}
F(X)=\left( f\left(x^1\right)^\textsc{t}, \dots, f\left(x^N\right)^\textsc{t} \right)^\textsc{t}
\end{equation}

\vspace{-0.3cm}
be the vector field which describes the reaction dynamics of all compartments, with the reaction kinetics $f(x^j)=- Z B \mathcal{K}(x^*) B^\textsc{t} \mathrm{Exp} \left(Z^\textsc{t} \mathrm{Ln}\left(\frac{x^j}{x^*}\right)\right)$, $j=1,\dots, N$.

The \emph{\textbf{open compartmental model}} of the reaction-diffusion system (\ref{eq:pHRDvec-calRel3Rd}) is given by
\begin{equation*}
\begin{split}
\dot X&=-\left( \left(*_0\right)^{-1} \!\otimes I_m\right) \!\!\bigg( \Delta_d \frac{X}{X^*} -\left( \mathbf{tr}\otimes I_m\right)^\textsc{t}\hat f_b\bigg) +F(X)\\
e_b&= \left(\mathbf{tr}\otimes I_m\right) \frac{X}{X^*}\,,
\end{split}
\end{equation*}
where $\otimes$ represents the Kronecker product, $I_m$ is the identity matrix of dimension $m\times m$, $R_d(X)\geq \alpha I_{m N_e}$, $\alpha >0$, and $N_e$ is the number of edges of the primal mesh. The Laplacian matrix of the simplicial complex is $\Delta_d=\left( \mathbf{d}\otimes I_m\right)^\textsc{t} \left(*_1 \otimes I_m\right) R_d(X) \left( \mathbf{d}\otimes I_m\right)$. Note that we have used $\mathbf{d}$ to denote $\mathbf{d}^0=-\left({\mathbf{d}}_\mathrm{i}^{n-1} \right)^\textsc{t}$ and $\frac{X}{X^*}=\left( \left(\frac{x^1}{x^*}\right)^\textsc{t},  \left(\frac{x^2}{x^*}\right)^\textsc{t}, \dots,  \left(\frac{x^N}{x^*}\right)^\textsc{t}\right)^\textsc{t}$.

The total energy of the system, the sum of energies of all compartments, is\vspace{-0.5cm} 
\[
G_d(X)=\sum_{j=1}^N G_i(x^j)V_{\sigma_j^0}\,,\vspace{-0.4cm}
\]
where $\sigma_j^0$ is the vertex corresponding to the state $x^j$ and $V_{\sigma_j^0}$ is the $n$-dimensional support volume obtained by taking the convex hull of the simplex $\sigma_j^0$ and 
and its dual cell $\star_\mathrm{i} \sigma_j^0$. Since $V_{\sigma_j^0}=|\sigma_j^0| |\star_\mathrm{i}\sigma_j^0|= |\star_\mathrm{i}\sigma_j^0|$, $j=1,\dots,N$, the total energy can be written as
\begin{equation}\label{eq:RDCompGd}
\begin{split}\vspace{-0.4cm}
\!\!G_d(X)\!=\!\sum_{j=1}^{N} G_j(x^j) |\star_\mathrm{i}\sigma_j^0|= \left( G_1,\dots, G_N\right)*_0\!\mathbf{1}_N.\!\!
\end{split}\vspace{-0.3cm}
\end{equation}
The distributed chemical potential as the gradient of (\ref{eq:RDCompGd}) is given as
\begin{equation}
\begin{split}\vspace{-0.4cm}
\frac{\partial G_d}{\partial X}=\begin{pmatrix}
      \frac{\partial G_1}{\partial x^1} |\star_\mathrm{i}\sigma_1^0|  \\
      \vdots\\
       \frac{\partial G_N}{\partial x^N} |\star_\mathrm{i}\sigma_N^0|  \\  
\end{pmatrix}=\left( *_0 \otimes I_m\right)\mathrm{Ln}\left(\frac{X}{X^*}\right).
\end{split}\vspace{-0.4cm}
\end{equation}

\vspace{-0.3cm}
\newsec{Compartmental Model}~Imposing zero-flux boundary conditions, $\hat f_b=0$, leads to the closed compartmental model\vspace{-0.2cm}
\begin{equation}\label{eq:ClosedCompartmentalRD}
\begin{split}\vspace{-0.3cm}
\!\!\!\!\dot X&\!=F^*(X)=\!-\left( \left(*_0\right)^{-1} \otimes I_m\!\right)  \Delta_d \frac{X}{X^*}  +F(X),\!\!
\end{split}\vspace{-0.3cm}
\end{equation}
with a positive initial condition $X(0)=X_0\in \mathbb{R}_+^{mN}$.

A simple but crucial observation is that $F_k^*(X)\geq0$ when $X_k=0$ for any $k=1,\ldots,mN$ and $X\in \bar{\mathbb{R}}_+^{mN}$. We have\vspace{-0.2cm}
\begin{equation*}
\begin{split}
&F_k^*(X) =- \underbrace{Z^{i} B \mathcal{K}(x^*) B^\textsc{t} \Exp\left(Z^\textsc{t} \Ln\left(\frac{x^j}{x^*}\right)\right)}_{\varsigma_R^{ij}}\\
& -\underbrace{|\star_\mathrm{i} \sigma_j^0|^{-1}  \left( \mathbf{d}\otimes I_m\right)_k^\textsc{t} \left(*_1 \otimes I_m\right) R_d(X) \left( \mathbf{d}\otimes I_m\right) \frac{X}{X^*}}_{\varsigma_D^{ij}},
\end{split}\vspace{-0.55cm}
\end{equation*}

\vspace{-0.7cm}
where $k=((j-1)m+i)$ and $X_k=x_i^j$, while $Z^i$ is the $i$-th row vector of $Z$ and $\left( \mathbf{d}\otimes I_m\right)_k^\textsc{t}$ is the $k$-th column of $\left( \mathbf{d}\otimes I_m\right)$. 

\vspace{-0.35cm}
When $X_k=x_i^j=0$, the terms corresponding to the positive $i$-th diagonal element of the weighted Laplacian matrix $B \mathcal{K}(x^*) B^\textsc{t}$ are all zero, while there is at least one term corresponding to a non-zero, and therefore strictly negative, off-diagonal element of $B \mathcal{K}(x^*) B^\textsc{t}$. This implies that $\varsigma_R^{ij}\leq0$. Similarly, the matrix $\left(*_1 \otimes I_m\right) R_d(X)$ is a diagonal strictly positive definite matrix, thus the terms corresponding to the positive $((j-1)m+i)$-th diagonal element of the augmented Laplacian matrix $\left( \mathbf{d}\otimes I_m\right)^\textsc{t}\! \!\left(*_1\! \otimes I_m\right) R_d(X) \left( \mathbf{d}\otimes I_m\right)$ are all zero, while there is at least one off-diagonal negative element. Thus, $\varsigma_D^{ij}\leq0$, and therefore $F^*(X)\geq 0$ when $X_k=0$.

\vspace{-0.3cm}
\begin{lem}
Suppose that $X:[0,t^*]\rightarrow \bar{\mathbb{R}}_+^{mN}$ is any solution of (\ref{eq:ClosedCompartmentalRD}). Then for any $k=1,\ldots,mN$:

\vspace{-0.7cm}
$$X_k(0)>0 ~~\Rightarrow~~ X_k(t^*)>0.$$
\end{lem}
\vspace{-0.6cm}
\begin{pf}
The proof is a repetition of the arguments given in \cite{Sontag}, Lemma 7.1, for a different class of systems.

\vspace{-0.3cm}
Suppose that $k$ is so that $X_k(0)>0$. Let $\Phi:\mathbb{R}^2\rightarrow \mathbb{R}$ be the function which for $t\in [0,t^*]$ and $y\in \mathbb{R}$ coincides with $\Phi(t,y)\!:=\!F_k^*\big(X_1(t),\ldots\!,\!X_{k-1}(t),y,X_{k+1}(t),\!\ldots\!,\!X_{mN}(t)\big)$ and has $\Phi(t,y)=\Phi(0,y)$ for $t<0$ and $\Phi(t,y)=\Phi(t^*,y)$ for $t>t^*$. Since $F_k^*(X)\geq 0$ when $X_k=0$, $\Phi(t,0)\geq 0$ for all $t$. For $t\in [0,t^*]$, the scalar function $y(t):=X_k(t)$ satisfies $\dot y(t)=\Phi(t,y(t))$. We need to prove that $y$ never vanishes. To this end, let $\Psi(t,p):=\Phi(t,p)-\Phi(t,0)$, and let $\dot z(t)=\Psi(t,z(t))$ with $z(0)=y(0)$.

\vspace{-0.2cm}
Because $\Psi$ is locally Lipschitz and $0$ is an equilibrium of $\dot z=\Phi(t,z)$, $z(t)>0$ for all $t$. Furthermore, $\dot z=\Psi (t,z)\leq \Phi (t,z)$ for all $t$, and thus by comparison $z(t)\leq y(t)$. Since $y(t^*)$ is well-defined, $z(t)$ remains bounded, and thus is defined for $t=t^*$. Hence, $y(t^*)\!\geq z(t^*)\!>0$.
\hfill $\qed$
\end{pf}

\vspace{-0.5cm}
\begin{cor}
For the system (\ref{eq:ClosedCompartmentalRD}) the positive orthant $\mathbb{R}_+^{mN}$ is forward invariant.
\end{cor}
\vspace{-0.2cm}

Jus like in the case of the system (\ref{masterequation}), in order to exclude the existence of possible boundary equilibria, we shall assume the global persistency property.
\vspace{-0.3cm}

\begin{conj}
Given $X_0\in \mathbb{R}_+^{mN}$, all the trajectories $t\mapsto X(t)$ of (\ref{eq:ClosedCompartmentalRD}) satisfy: $\mathrm{lim}\,\mathrm{inf}_{t\rightarrow \infty}X(t)>0$.
\end{conj}
\vspace{-0.2cm}

In the absence of the diffusion terms, the dynamics of the spatially discrete systems are decoupled, and as such coincide with the dynamics of the balanced reaction system (\ref{masterequation}). In this scenario all the compartments exhibit asymptotically stable dynamics, but the steady states of all the compartments, in general, are not identical. The following theorem shows that the compartmental model (\ref{eq:ClosedCompartmentalRD}) is asymptotically stable with the spatially uniform steady state. 

\vspace{-0.3cm}
\begin{thm}\label{tm:LyapRDcompartment}
Consider the compartmental model of balanced mass action reaction network given by (\ref{eq:ClosedCompartmentalRD}). For every initial condition $X(0)\in \mathbb{R}_+^{mN}$, the species concentrations $x^1, \dots, x^N$ as $t \to \infty$ converge to $x^1=\cdots=x^N\in\mathcal{E}$.
\end{thm}
\vspace{-0.6cm}
\begin{pf}
In \cite{AJReaction} the authors have shown that $G$ in (\ref{eq:FreeEnergyGibbs}) satisfies $G(x^{*}) =0$ and $ G(x) > 0$, $\forall x \neq x^{*}$, and
for every real $ c >0$ the set $\{x\in \bar{\mathbb{R}}_+^{m} \mid G(x) \leq c\}$ is compact. 
This easily can be checked. Let $x_i$ and $x_i^*$ denote the $i$-th elements of $x$ and $x^*$ respectively. From the strict concavity of the logarithmic function 
$ z- 1\geq \ln(z), $ $\forall z \in \mathbb{R}_+$, with equality if and only if $z=1$.
Putting $z=\frac{x_i^*}{x_i}$, we have $
x_i^*-x_i+x_i \ln \left(\frac{x_i}{x_i^*}\right) \geq 0
$, with equality if and only if $x_i = x_i^*$. This implies that
$
G(x)=\sum_{i=1}^m\left(x_i^*-x_i+x_i\ln\left(\frac{x_i}{x_i^*}\right)\right) \geq 0
$,
with equality if and only if $x_i = x_i^*, i=1, \cdots, m$. Thus $G$ has a strict minimum at $x=x^*$ and and $ G(x) > 0$, $\forall x \neq x^{*}$.

\vspace{-0.2cm}
The above stated properties of $G$ immediately imply that the function $X\mapsto G_d(X)$ in (\ref{eq:RDCompGd}) satisfies
\begin{equation}\label{lyap}
G_d(X^{*}) =0, \quad G_d(X) > 0 , \quad \forall X \neq X^{*},
\end{equation}
and is {\it proper}, i.e., for every real $ C >0$ the set $\{X\in \bar{\mathbb{R}}_+^{mN} \mid G_d(X) \leq C\}$ is compact. 

\vspace{-0.2cm}
In what follows we will show that $\dot{G}_d(X)= \frac{\partial^\textsc{t} G_d}{\partial X}(X) \dot X= \frac{d G_d}{dt}(X)$ satisfies
\begin{equation}\label{lyap1}
\dot{G_d}(X) \leq 0 \,\mbox{ for all }\,   X \in \mathbb{R}_+^{mN},
\end{equation}
and
\begin{equation}\label{lyap2}
\dot{G_d}(X) = 0\, \mbox{ if and only if }\, x^1=\cdots=x^N\in \mathcal{E}.
\end{equation}

We look for the time derivative of the total energy:
\begin{equation}
\begin{split}
\dot G_d&=\frac{\partial^\textsc{t} G_d}{\partial X} \dot X=\left( \left( *_0 \otimes I_m\right) \mathrm{Ln}\left(\frac{X}{X^*}\right)\right)^\textsc{t} \dot X\\
 & =-  \mathrm{Ln}\left(\frac{X}{X^*}\right)^\textsc{t} \left( *_0 \otimes I_m\right)^\textsc{t}  \left( \left(*_0\right)^{-1} \otimes I_m\right)\\
 &~~~~~  \cdot\left( \mathbf{d}\otimes I_m\right)^\textsc{t} \left(*_1 \otimes I_m\right) R_d(X)  \left( \mathbf{d}\otimes I_m\right) \frac{X}{X^*}  \\
 & ~~~ +  \mathrm{Ln}\left(\frac{X}{X^*}\right)^\textsc{t} \left( *_0 \otimes I_m\right)^\textsc{t} F(X).
\end{split}
\end{equation}
Since $\left( *_0 \otimes I_m\right)^\textsc{t}  \left( \left(*_0\right)^{-1}\otimes I_m\right)=I_{mN}$, we have
\begin{equation*}
\begin{split}
&\dot G_d\\
&=- \mathrm{Ln}\left(\frac{X}{X^*}\right)^\textsc{t} \!\!\left( \mathbf{d}\otimes I_m\right)^\textsc{t} \left(*_1 \otimes I_m\right) R_d(X) \left( \mathbf{d}\otimes I_m\right) \frac{X}{X^*}  \\
 & ~~~ +  \mathrm{Ln}\left(\frac{X}{X^*}\right)^\textsc{t} \left( *_0 \otimes I_m\right)^\textsc{t} F(X)\\
 &=\!-\! \left(\left( \mathbf{d}\otimes I_m\right)\mathrm{Ln}\!\left(\frac{X}{X^*}\!\right)\!\right)^\textsc{t} \!\!\! \left(*_1 \otimes I_m\right) R_d(X) \left( \mathbf{d}\otimes I_m\right)   \frac{X}{X^*}\\
 &~~~+ \sum_{i=1}^N |\star_\mathrm{i} \sigma_i^0| \underbrace{\frac{\partial^\textsc{t} G_i}{\partial x^i}f(x^i)}_{\varepsilon_R(x^i)}\\
  &= -\underbrace{\left\langle\left( \mathbf{d}\otimes I_m\right)\mathrm{Ln}\left(\frac{X}{X^*}\right) , R_d(X) \left( \mathbf{d}\otimes I_m\right) \frac{X}{X^*}  \right\rangle_d}_{\varepsilon_D }\\
 &~~~+ \sum_{i=1}^N |\star_\mathrm{i} \sigma_i^0| \varepsilon_R(x^i)
\end{split}
\end{equation*}

We compute the expression $\varepsilon_R(x^i)$, along the lines of \cite{AJReaction}, as
\begin{equation}\label{eq:epsilonR}
\begin{array}{rl}
\!\!\!\!\varepsilon_R(x^i) &= \frac{\partial^\textsc{t} G_i}{\partial x^i}f(x^i)\\
&= -\mu^\textsc{t}(x^i) ZB\mathcal{K}(x^*)B^\textsc{t} \Exp(Z^\textsc{t}\mu(x^i)) \\[2mm]
&= -\gamma^\textsc{t}(x^i) B\mathcal{K}(x^*)B^\textsc{t} \Exp(\gamma(x^i)) \\[2mm] 
&= \!\sum_{j=1}^r\!
\left(\gamma_{\mathcal{S}_j}(x^i)-\!\gamma_{\mathcal{P}_j}(x^i)\right)\kappa_j(x^*)\!\! \\&~~~~~~~~~~~\cdot \left(\exp\big(\gamma_{\mathcal{P}_j}(x^i) \big)\!-\exp\big(\gamma_{\mathcal{S}_j}(x^i) \big)\right)\!\!\!\!\!\\[2mm]
& \leq 0 \, ,
\end{array}
\end{equation}
since $\kappa_j(x^*) >0$ for $j=1,\ldots, r$, and the exponential function is strictly increasing. 
The summand in the third line of (\ref{eq:epsilonR}) is zero only if $\gamma_{\mathcal{S}_j}(x^i) - \gamma_{\mathcal{P}_j}(x^i) =0$ for every $j$. This is equivalent to having $B^\textsc{t}\gamma(x) = 0$. Thus, $\varepsilon_R(x^i)=0$ only if $B^\textsc{t}\gamma=B^\textsc{t}Z^\textsc{t}\Ln\left(\frac{x^i}{x^*}\right)=0$. It follows that
\vspace{-0.3cm}
\begin{equation}\label{eq:RDepR}
\begin{split}
\varepsilon_R(x^i)= 0\quad \mbox{if and only if}\quad x^i\in \mathcal{E}
\end{split}
\end{equation}
for all $\quad i=1,\dots, N$.

For the contribution of the compartmental diffusion dynamics we have
\begin{equation*}\label{eq:RDepsilonDiff}
\begin{split}
&\varepsilon_D\\
&=\left\langle\left( \mathbf{d}\otimes I_m\right)\frac{\partial G}{\partial X} , R_d(X) \left( \mathbf{d}\otimes I_m\right) \frac{X}{X^*}  \right\rangle_d\\
&\geq \sum_{k=1}^{N_e} \left( \mathrm{Ln}\left(\frac{x^i}{x^*}\right) -  \mathrm{Ln}\left(\frac{x^j}{x^*}\right) \right)^\textsc{t}\!\!\alpha |\sigma_k^1| \left( \frac{x^i}{x^*} -  \frac{x^j}{x^*}\right)\geq 0\,,
\end{split}
\end{equation*}
where $N_e$ is the number of edges of the primal mesh, and $x^i$ and $x^j$ are states associated to the nods $i$ and $j$, and $k$ is the edge between nods $i$ and $j$. Because $\mathrm{ln}(\cdot)$ is an increasing function, $\left( \mathrm{Ln}\left(\frac{x^i}{x^*}\right) -  \mathrm{Ln}\left(\frac{x^j}{x^*}\right) \right)$ possesses the same sign as $\left( \frac{x^i}{x^*} -  \frac{x^j}{x^*}\right)$, and hence $\varepsilon_D\geq 0$. Furthermore,
\begin{equation}\label{eq:RDepD}
\begin{split}
\varepsilon_D=0 \quad \mbox{if and only if}\quad x^1= \dots= x^N.
\end{split}
\end{equation}

Now, $\dot G_d=0$ if and only if $\varepsilon_R=0$ and $\varepsilon_D=0$. The intersection of the two conditions (\ref{eq:RDepD}) and (\ref{eq:RDepR}) gives (\ref{lyap2}).

Since $G_d$ is proper (in $\bar{{\mathbb{R}}}_+^{mN}$) and the state trajectory $t\mapsto X(t)$ remains in ${\mathbb{R}}_+^{mN}$, (\ref{lyap1}) implies that $t\mapsto X(t)$ is bounded in ${\mathbb{R}}_+^{mN}$. Therefore, boundedness of $t\mapsto X(t)$, together with equations (\ref{lyap1}) and (\ref{lyap2}), by LaSalle's invariance principle imply that all the species concentrations $x^1, \dots, x^N$ converge to an element in $\mathcal{E}$. \hfill $\qed$
\end{pf}

\begin{rmk}
Theorem \ref{tm:LyapRDcompartment} remains unaltered if we replace the reaction vector field (\ref{masterequation}) by any vector function of the form (not corresponding anymore to mass action kinetics)\vspace{-0.3cm}
\begin{equation}\label{eq:rmkMM}
f({x^j}) = - Z B \mathcal{K}_\mathrm{g}(x^j,x^*) B^\textsc{t} \Phi \left(Z^\textsc{t} \mathrm{Ln}\left(\frac{x^j}{x^*}\right)\right),
\end{equation}
where $\mathcal{K}_\mathrm{g}(x^j,x^*) := \mathrm{diag} \big( \kappa_1^\mathrm{g}(x^j,x^*), \cdots, \kappa_r^\mathrm{g}(x^j,x^*) \big)>0$ for all $x^j\in \mathbb{R}_+^m$ and $\Phi: \mathbb{R}^c \to \mathbb{R}^c$ is a mapping $\Phi(y_1,\cdots,y_c)= \diag (f_1(y_1), \dots,f_c(y_c))$, with the functions $\Phi_i$, $i=1,\dots,c$, all monotonically increasing.

\vspace{-0.3cm}
In fact, a recent paper \emph{{\cite{Shodhan}}} shows that, for instance, Michaelis-Menten kinetics are in the form (\ref{eq:rmkMM}), where $K_\mathrm{g}(x,x^*)$ is a rational but strictly positive definite matrix.
\end{rmk}

\section{Chemical Example}\label{Ch:RD:SectionExample}
We illustrate our analysis on a simple chemical reaction model
\begin{equation}\label{eq:RDexampleReaction}
\begin{split}\vspace{-0.4cm}
\xymatrix{
X_1 +X_2 \ar@^{->}@<0.5ex> ^-{k_1^\mathrm{forw}}[r] & \ar@^{->} @<0.5ex> [l]^-{k_1^\mathrm{rev}}X_3  \ar@^{->} @<0.5ex>[r]^-{k_2^\mathrm{forw}}& \ar@^{->} @<0.5ex> [l]^-{k_2^\mathrm{rev}}X_1 + X_4,
}\!\!\!
 \end{split}\vspace{-0.4cm}
\end{equation}
where $X_1$ is enzyme, $X_2$ substrate, $X_3$ intermediate product, and $X_4$ product. The first (binding) and third (unbinding) steps are reversible. Many reactions in the glycolysis metabolic pathway are of this type. 

\vspace{-0.2cm}
For instance, \emph{glucose-6-phosphate isomerase} (alternatively known as \emph{phosphoglucose isomerase} or \emph{phosphohexose isomerase}) is an enzyme that catalyzes the conversion of \emph{glucose 6-phosphate} (G6P) into \emph{fructose 6-phosphate} (F6P) in the second step of glycolysis. The change in structure is an isomerization, in which the G6P ($X_1$) has been converted to F6P ($X_4$). The freely reversible reaction requires an enzyme $X_2$, phosphohexose isomerase, to proceed; for more details see, e.g., \cite{Berg}.


\vspace{-0.2cm}
The dynamical model of (\ref{eq:RDexampleReaction}) governed by mass action kinetics is given by
\begin{equation}\label{eq:RDbalancedExample}
\begin{split}
\dot x_1 &= -k_1^\mathrm{forw} x_1 x_2+(k_2^\mathrm{forw}+k_1^\mathrm{rev}) x_3-k_2^\mathrm{rev} x_1 x_4\\
\dot x_2 &=    -k_1^\mathrm{forw} x_1 x_2+k_1^\mathrm{rev}x_3\\
\dot x_3 &=    k_1^\mathrm{forw} x_1 x_2-(k_1^\mathrm{rev}+k_2^\mathrm{forw}) x_3+k_2^\mathrm{rev} x_1 x_4\\
 \dot x_4 &=   k_2^\mathrm{forw} x_3-k_2^\mathrm{rev} x_1 x_4\,.
 \end{split}
\end{equation}

\vspace{-0.3cm}
It can be easily checked that $x_1^*=x_2^*=1$, $x_3^*=k_1^\mathrm{forw}/k_1^\mathrm{rev}$, $x_4^*=k_1^\mathrm{forw} k_2^\mathrm{forw} / (k_1^\mathrm{rev}k_2^\mathrm{rev})$ is one of the equilibria of the system (\ref{eq:RDbalancedExample}). The complex stoichiometric matrix $Z$, the incidence matrix $B$, and the stoichiometrric matrix $S$ for the reaction network (\ref{eq:RDexampleReaction}) are
\begin{equation*}
\begin{split}
Z\!=\!\begin{pmatrix}
      1 & 0 & 1   \\
     1 & 0 & 0\\
     0 & 1& 0\\
     0 & 0 &1 
\end{pmatrix}\!,~
B\!=\!\begin{pmatrix}
      1 & 0   \\
     -1 & 1\\
     0 & -1
\end{pmatrix}\!, ~S\!=\!ZB=\!\begin{pmatrix}
      -1 & 1   \\
     -1 & 0\\
     1 & -1\\
     0 & 1 
\end{pmatrix}\!.
 \end{split}
\end{equation*}
Since $Z_{\mathcal{S}_1}=\left(1,1,0,0\right)^\textsc{t}$, $Z_{\mathcal{P}_1}=Z_{\mathcal{S}_2}=\left(0,0,1,0\right)^\textsc{t}$, and $Z_{\mathcal{P}_2}=\left(1,0,0,1\right)^\textsc{t}$, for the chosen $x^*$, the diagonal balanced reaction constants (cf. Section~\ref{Sec2}) are
\begin{equation*}
\begin{split}
\mathcal{K}(x^*)=\begin{pmatrix}
      k_1^\mathrm{forw} & 0\\
      0 & \frac{k_1^\mathrm{forw}k_2^\mathrm{forw}}{k_1^\mathrm{rev}}
\end{pmatrix}.
 \end{split}
\end{equation*}
The system (\ref{eq:RDbalancedExample}) now can be rewritten into the form (\ref{masterequation}), while the dynamics under the influence of diffusion is given by the reaction-diffusion model (\ref{eq:pHRDvec-calRel3Rd}).

\begin{figure}
 \hspace{-0.0cm}\includegraphics[width=8.3cm,angle=0]{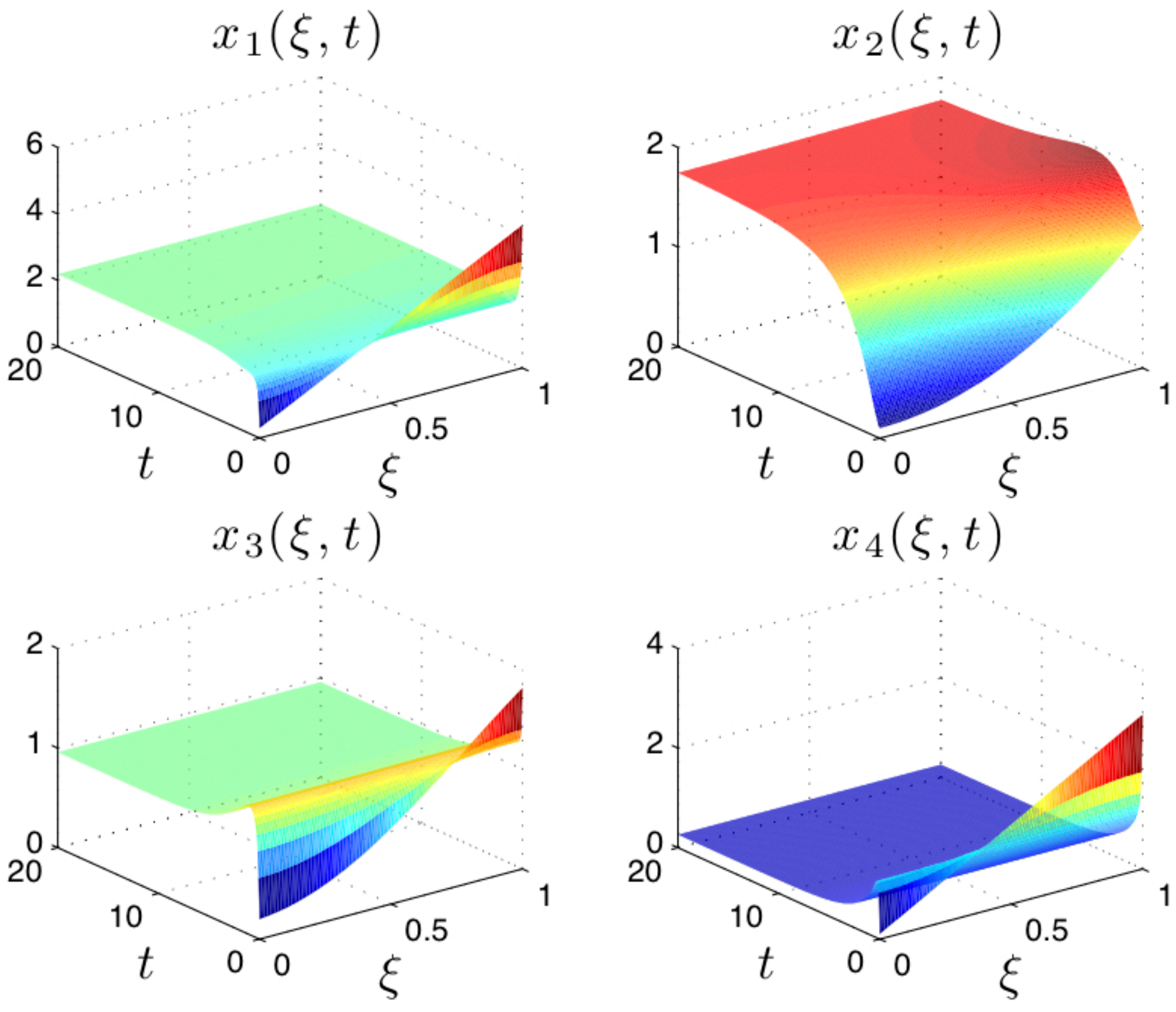}\\
  \caption{Solutions of (\ref{eq:RDbalancedExample}), in the presence of the diagonal diffusion term $\mathrm{diag}(d_1,d_2,d_3,d_4)\Delta x$, on the one-dimensional spatial domain $M=[0,1]$ with initial conditions $x_1(\xi,0)=4\xi+0.3$, $x_2(\xi,0)=1.3\xi^2+0.1$, $x_3(\xi,0)=2\sin^2(\xi)+0.2\xi+0.2$, $x_4(\xi,0)=3\xi+0.1$, and Neumann's boundary conditions. Diffusion coefficients are set to be $d_1=0.33$, $d_2=0.72$, $d_3=0.91$, and $d_4=0.67$. The reaction rates are $k_{1}^\mathrm{forw}=0.1$, $k_{1}^\mathrm{rev}=0.4$, $k_{2}^\mathrm{forw}=0.3$, $k_{2}^\mathrm{rev}=0.5$. Upon the transient phase the system reaches a steady state $ x^{**}=x(\xi,\infty)=(2.1856, 1.7557, 0.9602, 0.2638)^\textsc{t}$ uniform in space. Immediately, we verify that $x^{**}$ is a thermodynamical equilibrium, $S^\textsc{t}\mathrm{Ln}(x^{**})=S^\textsc{t}\mathrm{Ln}(x^{*})$. The number of compartments used in $N=20$.}\label{fig:RD4plotDneq0}
\end{figure}

The spatially uniform asymptotic behavior predicted by Theorem~\ref{tm:LyapRDcompartment} is demonstrated with the simulation in Figure~\ref{fig:RD4plotDneq0}. The elements of the matrix $R_d$ in the system (\ref{eq:ClosedCompartmentalRD}) given in terms of the standard diffusion matrix $D$ are $R_d:=\big(\mathrm{diag}(\frac{1}{x_1^*},\ldots, \frac{1}{x_m^*})D\big)\otimes I_m$, where in our case $m=4$. Eliminating the effects of diffusion leads to a spatially nonuniform steady state.

%

 \section{Concluding Observations}
We have provided a geometric formulation of reaction-diffusion systems with a thermodynamical equilibrium. Diffusion as an isolated process is associated with a homogenizing effect that eliminates the gradients of the constituents and eventually leads to uniform spatial state. However, diffusion in combination with reaction dynamics can produce spatially heterogenous patterns. We envision that the Hamiltonian perspective presented in this paper will utilize the systems analysis of reaction-diffusion system and foster new insights in compartmental systems design.


\vspace{-0.2cm}
Control for reaction-diffusion systems in the port-Hamiltonian framework can be understood as the coupling of a reaction-diffusion system to an additional port-Hamiltonian system that plays the role of the controller. This, among others, enables the application of passivity base techniques in control synthesis for reaction-diffusion systems. Only by having accurate structured discretization can one hope to approach this challenging enterprise.

\vspace{-0.2cm}
A model obtained by structure-preserving scheme for the spatial discretization is a compartmental model, which exhibits a striking similarity with consensus dynamics \cite{Murray,Cortes}. Exploring this resemblance is a very appealing research direction.

\vspace{-0.3cm}

\end{document}